\documentclass
[10pt,a4paper,tightenlines,twocolumn,prd,showpacs,showkeys,lengthcheck,nobibnotes]{revtex4}%

\usepackage{amsfonts}
\usepackage{amsmath}
\usepackage{amssymb}
\usepackage{epsfig}
\usepackage{graphicx}
\usepackage{fancyhdr}%
\begin{document}
\preprint{ }
\title[ ]{\textbf{Quantized detector network POVMs and the Franson-Bell experiment}}
\author{George Jaroszkiewicz}
\affiliation{University of Nottingham, University Park, Nottingham, NG7 2RD, UK}
\keywords{quantized detector networks, POVMs, Franson-Bell experiment}
\pacs{03.65.Ca, 03.65.Ud, 03.65.-a}

\begin{abstract}
We present a generalized POVM formalism for the calculation of quantum optics
networks of arbitrary complexity and apply it in a detailed analysis of the
Franson-Bell experiment. Our analysis suggests that when it comes to observations of quantum processes, laboratory apparatus cannot always be used according to classical principles.

\end{abstract}
\maketitle

\section{Introduction}

The mathematical rules of standard quantum mechanics (SQM) are now well
understood, but for over a hundred years it has been a serious challenge to
understand what quantum mechanics (QM) means on an intuitive level. Part of
the problem has been a natural tendency to regard QM as a direct replacement
for classical mechanics (CM). This led to the belief that SQM is the most
complete and correct description possible of systems under observation (SUOs)
such as electrons, photons and atoms. This belief was challenged in the EPR
paper \cite{EPR-1935} but SQM\ won the argument because to date all of its
predictions have been supported empirically, including those for Bell
inequality experiments. As a result, many theorists readily use phrases such
as \emph{electron wave-function, photon polarization state} and \emph{atomic
orbital}, as if these had some sort of physical existence independent of any context.

This view is understandable given the success of the Schr\"{o}dinger equation
when applied to important SUOs such as atoms and molecules. Some of the
consequences of taking the Schr\"{o}dinger equation literally were the
development of Everett's relative state formulation \cite{EVERETT-1957} and
decoherence theory \cite{ZUREK-2002}.

Whilst these developments contain much of interest, particularly in the case
of decoherence theory, the finger of empirical evidence points firmly in
another direction, towards the fundamental importance of \emph{apparatus} in
the interpretation of QM. Heisenberg and Bohr understood from the beginning
that SUOs and apparatus cannot be separated in QM in the way that they are in
CM. Indeed, Bohr's counterattack to the EPR paper \cite{BOHR-1935} relied on
the contextuality of the processes of observation.

In recent times this point has been strongly reinforced by theorists such as
Ludwig \cite{LUDWIG-1983A} and Kraus \cite{KRAUS-1983}. The work of these and
others such as Peres \cite{PERES:1993} has supported the view that QM\ is much
more about information exchange between SUOs and apparatus rather than being
just an improved description of SUOs. Given this, it is natural to ask who or
what is acquiring quantum information and how it happens, questions which
focus attention on observers and apparatus. This was the motivation for this paper.

Whilst the rules of SUOs are those given by SQM and are therefore well
understood mathematically, the quantum rules as they relate to apparatus
remain to be fully understood and developed. There are surprises, as we shall
show in our description of the Franson-Bell experiment, discussed in detail below.

In an effort to understand the processes of observation and measurement in QM,
we have in recent years concentrated our efforts in developing \emph{quantized
detector networks} (QDN). This is a time-dependent description of apparatus
based on quantum registers \cite{J2007C}. In this paper we combine QDN with
the SQM rules of SUOs and the Ludwig-Kraus POVM formalism to provide a general
framework for a wide class of modularized quantum processes.

In recent years, quantum optics has provided an ideal arena for the
theoretical and experimental investigation of quantum principles, principally
because photonics is a clean technology amenable to a modular approach. By
this we mean that experimental apparatus of great complexity can be built up
from basic components such as beam-splitters, Wollaston prisms, mirrors,
polarizers and suchlike. Experiments based on such modules are the targets of
our work described here.

In this paper we have two principal objectives. The first is to outline the
generalized POVM formalism we have developed to describe modular quantum
optical networks of arbitrary complexity. This formalism combines the two
complementary aspects fundamental to all quantum experiments, viz., SUOs and
apparatus. We have found this formalism well suited to algebraic
computerization. All our reported POVM elements and coincidence rates have
been calculated using a single computer algebra program, details of which can
be supplied on request. This approach appears capable of dealing with modular
networks of arbitrary complexity.

The second objective is to apply our formalism to a particular experiment
first discussed theoretically by Franson \cite{FRANSON-1989}\ and referred to
here as the Franson-Bell experiment. This is of considerable interest
empirically because it appears to involve quantum non-locality and Bell-type
violations of classical expectations. Our network approach to the Franson-Bell
experiment shows that \emph{laboratory apparatus} \emph{cannot under all
circumstances be treated classically}. Specifically, our analysis shows that
how a photon detector's dynamical output is determined depends on the
observer's knowledge of the photon states concerned. Another way of saying
this is that the interpretation of what real equipment in a laboratory means
is contextual and depends on the input states involved.

\section{Notation, terminology and rules}

A quantized detector network is a time-dependent collection of \emph{quantum
nodes} and \emph{quantum paths}. Quantum nodes are those places in a quantized
network at which an elementary \emph{yes/no} observation is possible in
principle. They are equivalent to \emph{effects} in the Ludwig-Kraus formalism
\cite{KRAUS-1983} and to elementary signal detectors (ESDs) in QDN
\cite{J2007C}. They can serve as sources and detectors of quantum signals.
Quantum path are those parts of a quantized network connecting two nodes. The
quantum optics modules we discuss are always placed on quantum paths, i.e.,
between the quantum nodes where the detectors sit.

Time is treated with particular care in our formalism. Our notion of time is
closer to proper time in relativity than to coordinate time and is a measure
of information acquisition by observers, rather than a parameter dictating
changes in SUOs. It always runs forwards and is irreversible. We assume from
the outset that quantum experiments are dynamical processes in which not only
states of SUOs evolve but apparatus itself can change.

In our approach, an observer extracts information during any given run of an
experiment in a sequence of \emph{stages }\cite{J2005A}. Usually a stage
$\Omega_{n}$ can be identified as some instant of simultaneity of the
observer's laboratory time, but this is not always the case. A stage can
involve parts of an experiment $``$on hold$"$, i.e., isolated from all other
parts in such a way that to all intents and purposes no dynamical interaction
or information flow can take place between them. A stage is a collection of
detectors which are all \emph{effectively relatively spacelike}, even though
their laboratory times may indicate that some of them are relatively timelike.
We have used this $``$multi-fingered$"$ concept of time in a discussion of
particle decay experiments and the quantum Zeno effect \cite{J2007G}. The
Franson-Bell experiment provides an example where an intermediate stage in an
experiment is not synonymous with a single instant of the observer's
laboratory time.

One of the consequences of our approach is that in our formalism, quantum
dynamics is always described in terms of mappings from one Hilbert space to
another. There is no concept here of a state vector evolving in a fixed
Hilbert space. Therefore, we are not dealing with the Schr\"{o}dinger picture
as we would normally in SQM. Also, because a jump from one Hilbert space to
another may involve a change of Hilbert space dimension, we are not dealing
with the Heisenberg picture either. In our diagrams, a symbol such as
$A_{n}^{i}$ denotes the $i^{th}$ detector associated with stage\emph{\ }%
$\Omega_{n}$. Note that $A_{n}^{i}$ need bear no causal relation to $A_{m}%
^{i}$ whatsoever, if $n\neq m$.

In all the discussions here, we shall concentrate exclusively on pure state
physics. This is not only because mixed states are not expected to provide any
great difficulties for the formalism, but also because the lessons to be
learnt from the Franson-Bell experiment come from the fundamental level of
pure state physics. We shall ignore all effects of quantum inefficiency and
probability leakage, etc., as these bear the same relationship to our theory
as friction does in CM, i.e., they are important aspects but do not contribute
to the discussion.

In a typical experiment, we shall deal with tensor product states of the form
\begin{equation}
\Psi_{n}\equiv\sum_{i=1}^{d_{n}}\sum_{A=0}^{D_{n}-1}\Psi_{n}^{iA}s_{n}%
^{i}a_{n}^{A},\label{POVM-01}%
\end{equation}
where $s$ refers to the SUO part, $a$ refers to the apparatus and the
coefficients $\Psi_{n}^{iA}$ are complex numbers. From the point of view of
decoherence theory, $a$ would normally be replaced by $e$, to denote
\emph{environment}. However, there are crucial differences between that theory
and our approach. Specifically, we do not use the Schr\"{o}dinger equation to
describe apparatus. Moreover, we do not exclude the state reduction concept,
as this is associated with information exchange between observer and apparatus
and is not interpreted by us as the collapse of any wave-like object.

In an expression such as (\ref{POVM-01}) above, $s_{n}^{i}$ denotes a basis
ket state $|s^{i},n\rangle$ in some separable Hilbert space $\mathcal{H}_{n}$
of dimension $d_{n}$, with dual space $\mathcal{\bar{H}}_{n}$. Generally, we
use small Latin letters to label SUO basis states and such an index runs from
$1$ to $d_{n}$. The dual $\langle s^{i},n|$ of $s_{n}^{i}$ is denoted by
$\bar{s}_{n}^{i}$.\ The $s_{n}^{i}$ are orthonormalized, so we have the
relation $\bar{s}_{n}^{i}s_{n}^{j}=\delta_{ij}$. The ordering of symbols
matters here, because $s_{n}^{j}\bar{s}_{n}^{i}$ denotes the operator
$|s^{j},n\rangle\langle s^{i},n|$. In our computer algebra program, ordering
can be ignored, provided inner products are always immediately evaluated.

On the other hand, the apparatus basis states (referred to as labstates)
$a_{n}^{A}$ are elements of a quantum register $\mathcal{R}_{n}$ which is the
tensor product of a finite number $r_{n}$ of qubits. $\mathcal{R}_{n}$ is a
Hilbert space with dimension $D_{n}=2^{r_{n}}$. Generally we shall use capital
Latin letters to index basis labstates and such labels will run from zero to
$D_{n}-1$. Full details are given in a recent review \cite{J2007C}. Typically,
a preferred basis element $a_{n}^{A}$ for $\mathcal{R}_{n}$ involves a number
$k$ of detectors combined at stage $\Omega_{n}$ in the form%
\begin{equation}
a_{n}^{M}\equiv\mathbb{A}_{m_{1},n}^{+}\mathbb{A}_{m_{2},n}^{+}\ldots
\mathbb{A}_{m_{k},n}^{+}|0,n),\label{POVM-05}%
\end{equation}
where $M=2^{m_{1}-1}+2^{m_{2}-1}+\ldots+2^{m_{k}-1}$, $\mathbb{A}_{m,n}^{+}$
is the signal operator for the $m^{th}$ detector at stage $\Omega_{n}$ and
$|0,n)$ is the \emph{void state} for stage $\Omega_{n}$. Here the integers
$m_{1}$, $m_{2}$, $\ldots$, $m_{k}$ are all necessarily different. We shall
also use the notation $a_{n}^{M}\equiv A_{n}^{m_{1}}A_{n}^{m_{1}}\ldots
A_{n}^{m_{k}}$.

The reason for the relative simplicity of our computer algebra approach is
that whilst the physics of a network experiment can involve multiple
detectors, such as in coincidence experiments, rule (\ref{POVM-05}) associates
a unique label to each combination of detectors, no matter how complicated.
This permits relatively straightforward programming.

Our strategy is not to provide a completely comprehensive description of SUOs
states, as would happen in a quantum field theory approach, but to focus on
and utilize \emph{observational context}. By this we mean that the intentions
of the observer dictate the choice of SUO space $\mathcal{H}_{n},$ and it is
this choice which permits the apparatus to be described relatively simply via
a quantum register $\mathcal{R}_{n}$. Essentially, the observer will have
decided in advance precisely which dynamical variables their apparatus is
registering yes/no answers for, and also which variables are not being
observed and are therefore irrelevant to the experiment. For instance, a
Stern-Gerlach apparatus provides information about electron spin but not about
electron momentum.

\section{Dynamical Evolution}

In the following, all Hilbert spaces involved are complex and finite
dimensional. Our quantum dynamics will be described in terms of probability
conserving linear maps from one total Hilbert space $\mathcal{H}_{n}%
\otimes\mathcal{R}_{n}$ to the next, $\mathcal{H}_{n+1}\otimes\mathcal{R}%
_{n+1}$. To understand the details of the calculations, we need a few basic concepts.

We define a \emph{Born map} to be a norm-preserving map from one Hilbert space
$\mathcal{H}$ into another Hilbert space $\mathcal{H}^{\prime}$. If
$\mathfrak{B:}\mathcal{H}\rightarrow\mathcal{H}^{\prime}$ is a Born map, then
for any element $\Psi$ in $\mathcal{H}$, the corresponding element
$\Psi^{\prime}\equiv\mathfrak{B}(\Psi)$ in $\mathcal{H}^{\prime}$ is such that
$(\Psi,\Psi)=(\Psi^{\prime},\Psi^{\prime})^{\prime},$where the inner product
on the LHS is taken in $\mathcal{H}$ whilst that on the RHS is taken in
$\mathcal{H}^{\prime}$. From the basic definition of a Hilbert space, only the
zero vector in $\mathcal{H}$ can be mapped into the zero vector in
$\mathcal{H}^{\prime}$ by a Born map.

Born maps can be non-linear. A \emph{semi-unitary} map is just a linear Born
map. Because linearity is fundamental to many quantum processes (but not all),
we shall focus on such maps exclusively throughout this paper. The following
theorem is fundamental to our work and is easy to prove:

\noindent\textbf{Theorem}: \emph{A semi-unitary map from Hilbert space
}$\mathcal{H}$\emph{ into Hilbert space }$\mathcal{H}^{\prime}$\emph{ exists
if and only if }$\dim\mathcal{H}^{\prime}\geqslant\dim\mathcal{H}$\emph{.}

If $U$ is a semi-unitary operator from $\mathcal{H}$ to $\mathcal{H}^{\prime}$
then $U^{+}$ is a map from $\mathcal{H}^{\prime}$ to $\mathcal{H}$ such that
$U^{+}U=I$, the identity operator for $\mathcal{H}$. This means that a
semi-unitary map not only preserves norms but also inner products, which turns
out useful in the construction of the dynamics.

If $\dim\mathcal{H}^{\prime}=\dim\mathcal{H}$, then $U^{+}$ is a semi-unitary
map from $\mathcal{H}^{\prime}$ to $\mathcal{H}$ such that $UU^{+}=I^{\prime}%
$, the identity operator for $\mathcal{H}^{\prime}$. In this special case, $U$
is said to be \emph{unitary}. If however $\dim\mathcal{H}^{\prime}%
>\dim\mathcal{H}$, then necessarily $UU^{+}\neq I^{\prime}$. Particle decay
experiments are examples of situations where the quantum evolution is
semi-unitary but not unitary \cite{J2007G}.

The dynamics underlying a given experiment is treated stage by stage, with
semi-unitary evolution from an initial state $\Psi_{0}\equiv\sum_{i=1}^{d_{0}%
}\sum\nolimits_{A=0}^{D_{0}-1}\Psi_{0}^{iA}s_{0}^{i}a_{0}^{A}$ to some final
state $\Psi_{N}\equiv\sum_{i=1}^{d_{N}}\sum\nolimits_{A=0}^{D_{N}-1}\Psi
_{N}^{iA}s_{N}^{i}a_{N}^{A}$, for $N>0$. The dimension $d_{n}$ of the SUO
Hilbert space $\mathcal{H}_{n}$ can change from one stage to another, as
happens in the case of parametric down conversion production of photon pairs
from a single photon.

Consider evolution from stage $\Omega_{n}$ to $\Omega_{n+1}$. In our
experiments, the observer is assumed to know which initial basis states need
to be considered, because such contextual information comes with the
construction of the apparatus. Typically there will be very many potential
quantum register basis labstates, but most of these will be irrelevant in a
given experiment. This occurs for example in experiments where photons come in
from multiple correlated sources. Therefore, we need concentrate only on those
basis states which are needed. These form a basis $\mathsf{B}_{n}^{eff} $ for
some effective subspace $\mathcal{H}_{n}^{eff}$ of the total space
$\mathcal{H}_{n}\otimes\mathcal{R}_{n}$. Provided we take all of the basis
elements of $\mathcal{H}_{n}^{eff}$ into account, we can restrict ourselves to
that subspace in our calculations. Typically, for such an element $s_{n}%
^{i}a_{n}^{A}$ in $\mathsf{B}_{n}^{eff}$, we write%
\begin{equation}
U_{n+1,n}s_{n}^{i}a_{n}^{A}=\sum\limits_{j=1}^{d_{n+1}}\sum\limits_{B=0}%
^{D_{n+1}-1}U_{n+1,n}^{jB,iA}s_{n+1}^{j}a_{n+1}^{B},
\end{equation}
where the complex coefficients $U_{n+1,n}^{jB,iA}$ satisfy the semi-unitarity
conditions%
\begin{equation}
\sum\limits_{j=1}^{d_{n+1}}\sum\limits_{B=0}^{D_{n+1}-1}[U_{n+1,n}%
^{jB,iA}]^{\ast}U_{n+1,n}^{jB,kC}=\delta_{ik}\delta_{AC},\ \ \ \ \
\end{equation}
and are determined from an empirical knowledge of the modules involved in the
network (analogous to specifying a Hamiltonian in SQM). Provided we restrict
the action of $U_{n+1,n}$ to $\mathcal{H}_{n}^{eff}$, we may use $``
$effective completeness$"$, i.e., resolution of the identity $I_{n}^{eff}$ in
$\mathcal{H}_{n}^{eff}$, to write%
\begin{equation}
U_{n+1,n}\backsimeq\sum\limits_{j=1}^{d_{n+1}}\sum\limits_{B=0}^{D_{n+1}%
-1}\sum\limits_{i}^{eff}\sum\limits_{A}^{Eff}s_{n+1}^{j}a_{n+1}^{B}%
U_{n+1,n}^{jB,iA}\bar{s}_{n}^{i}\bar{a}_{n}^{A},
\end{equation}
where we use the symbol $\backsimeq$ to denote effective representation and
the summations for $\Omega_{n}$ are over the required effective ranges. Then
semi-unitarity is equivalent to%
\begin{equation}
U_{n+1,n}^{+}U_{n+1,n}\backsimeq I_{n}^{eff}.
\end{equation}

The complete effective evolution operator $U_{N,0}$ from $\Omega_{0}$ to
$\Omega_{N}$ is given by $U_{N,0}=U_{N,N-1}U_{N-1,N-2}\ldots U_{2,1}U_{1,0}$
and is of the form%
\begin{equation}
U_{N,0}\backsimeq\sum\limits_{i=1}^{d_{N}}\sum\limits_{A=0}^{D_{N}-1}%
\sum\limits_{j}^{eff}\sum\limits_{B}^{Eff}s_{N}^{i}a_{N}^{A}U_{N,0}%
^{iA,jB}\bar{s}_{0}^{j}\bar{a}_{0}^{B},
\end{equation}
where the $U_{N,0}^{iA,jB}$ are given by a product of all the interstage
transition matrices involved. $U_{N,0}$ is semi-unitary provided each of the
interstage evolution operators $U_{n+1,n}$ is semi-unitary.

Once the complete evolution operator has been determined, the generalized POVM
operators are obtained in two steps. First we calculate the $D_{N}$
generalized Kraus operators $M_{N,0}^{A}$ by the rule%
\begin{align}
M_{N,0}^{A}  & =\bar{a}_{N}^{A}U_{N,0}\nonumber\\
& \backsimeq\sum\limits_{i=1}^{d_{N}}\sum\limits_{j}^{eff}\sum\limits_{B}%
^{Eff}s_{N}^{i}U_{N,0}^{iA,jB}\bar{s}_{0}^{j}\bar{a}_{0}^{B}%
,\ \ \ A=0,1,\ldots,D_{N}-1.
\end{align}
For most networks, particularly those with large rank, most of these operators
turn out to be zero. The $D_{n}$ elements $E_{N,0}^{A}$ of the POVM are then
calculated by the rule%
\begin{equation}
E_{N,0}^{A}\equiv M_{N,0}^{A+}M_{N,0}^{A}=\sum\limits_{i,j,k}^{eff}%
\sum\limits_{B,C}^{Eff}s_{0}^{k}a_{0}^{C}[U_{N,0}^{iA,kC}]^{\ast}%
U_{N,0}^{iA,jB}\bar{s}_{0}^{j}\bar{a}_{0}^{B}.
\end{equation}
They are positive operators over the initial total space $\mathcal{H}%
_{0}^{eff}$ and satisfy the rule%
\begin{equation}
\sum_{A=0}^{D_{N}-1}E_{N,0}^{A}=I_{0}^{eff}.
\end{equation}
In this respect they differ from the standard POVM formalism, the POVMs of
which are operators on the SUO Hilbert space $\mathcal{H}_{0}$.

Given a properly normalized initial state $\Psi_{0}\equiv\sum\limits_{i}%
^{eff}\sum\limits_{A}^{Eff}\Psi_{0}^{iA}s_{0}^{i}a_{0}^{A}$, then the outcome
probability rates at stage $N$ are given by%
\begin{equation}
\Pr(a_{N}^{A}|\Psi_{0})=\bar{\Psi}_{0}E_{N,0}^{A}\Psi_{0},\label{POVM-04}%
\end{equation}
where $\bar{\Psi}_{0}\equiv\sum\limits_{i}^{eff}\sum\limits_{A}^{Eff}[\Psi
_{0}^{iA}]^{\ast}\bar{s}_{0}^{i}\bar{a}_{0}^{A}.$

Care has to be taken in interpreting these quantities because QDN networks are
time dependent and also because incoming states have finite wave-trains (or
coherence times). In situations where wavetrains coming from different parts
of a network fail to overlap in time at a detector, no interference can be
expected to occur normally. This happens in the Franson-Bell experiment
Scenario $ii)$ discussed below. Note however that the Franson-Bell experiment
Scenario $iii)$ provides a spectacular exception to this rule.

Our formalism is comprehensive, in that the computer algebra programme
automatically generates single detector outcome rates $\Pr(A_{N}^{i}|\Psi
_{0})$, two photon coincidence rates $\Pr(A_{N}^{i}A_{n}^{j}|\Psi_{0})$, and
all possible higher order multiple detector incidence rates, up to the
saturation coincidence rate $\Pr(A_{N}^{1}A_{n}^{2}\ldots A_{N}^{r_{N}}%
|\Psi_{0})$. Most higher order incidence detector rates will be zero in
typical experiments involving a single photon or a single photon pair process.

In the next three sections we show how to apply this formalism to three test
cases: the Wollaston prism, the non-polarizing beam-splitter, and a more
complicated network discussed by Brandt. Then we discuss the Franson-Bell
experiment is detail.

\section{The Wollaston prism}

A Wollaston prism is a quantum optics module which splits up a beam of light
into two orthogonally polarized beams. Figure $1$ is a schematic diagram of
such a device. Here, symbols such as $A_{n}^{i}$ represent the $i^{th}$
detector in stage $\Omega_{n}$.

Consider an initial state $\Psi_{0}^{1}=(\alpha s_{0}^{1}+\beta s_{0}%
^{2})a_{0}^{1}\equiv\{\alpha|s^{1},0\rangle+\beta|s^{2},0\rangle
\}\otimes\mathbb{A}_{1,0}^{+}|0,0)$, where $s_{0}^{i}\equiv|s^{i},0\rangle$ is
a polarization state vector in two-dimensional photon polarization Hilbert
space $\mathcal{H}_{0}$, with $i=1$ or $2$ representing orthogonal
polarizations such as horizontal and vertical. Here the coefficients $\alpha$
and $\beta$ are complex and we assume normalization to unity, viz.,
$\left\vert \alpha\right\vert ^{2}+\left\vert \beta\right\vert ^{2}=1$.

\begin{figure}[h]
\centerline{\psfig{file=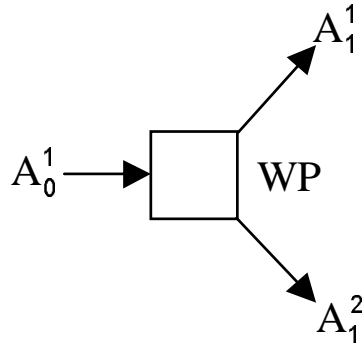,width=2in}}
\caption{The Wollaston prism.}%
\end{figure}

The evolution operator $U_{10}$ from stage $\Omega_{0}$ to stage $\Omega_{1}$
is assumed semi-unitary, mapping from a two-dimensional effective subspace of
the four-dimensional initial Hilbert space $\mathcal{H}_{0}\otimes
\mathcal{R}_{0}\ $to the final eight dimensional one $\mathcal{H}_{1}%
\otimes\mathcal{R}_{1}$. The rules are%
\begin{equation}
U_{10}s_{0}^{1}a_{0}^{1}=s_{1}^{1}a_{1}^{1},\ \ \ U_{1,0}s_{0}^{2}a_{0}%
^{1}=s_{1}^{2}a_{1}^{2},
\end{equation}
where $s_{1}^{i}a_{1}^{2^{j-1}}\equiv|s^{i},1\rangle\otimes\mathbb{A}%
_{j,1}^{+}|0,1).$ Using effective completeness we may write%
\begin{equation}
U_{10}\backsimeq s_{1}^{1}\bar{s}_{0}^{1}a_{1}^{1}\bar{a}_{0}^{1}+s_{1}%
^{2}\bar{s}_{0}^{2}a_{1}^{2}\bar{a}_{0}^{1},\label{POVM-02}%
\end{equation}
which satisfies the semi-unitary relation
\begin{equation}
U_{10}^{+}U_{10}=\{s_{0}^{1}\bar{s}_{0}^{1}+s_{0}^{2}\bar{s}_{0}^{2}%
\}a_{0}^{1}\bar{a}_{0}^{1}=I_{0}^{eff}.
\end{equation}

There are two non-zero generalized Kraus matrices associated with stage
$\Omega_{1}$, given by%
\begin{align}
M_{10}^{1}  & \equiv\frac{\partial}{\partial a_{1}^{1}}U_{1,0}=s_{1}^{1}%
\bar{s}_{0}^{1}\bar{a}_{0}^{1},\nonumber\\
M_{10}^{2}  & \equiv\frac{\partial}{\partial a_{1}^{2}}U_{1,0}=s_{1}^{2}%
\bar{s}_{0}^{2}\bar{a}_{0}^{1}.
\end{align}
From these, the generalized POVM operators associated with $\Omega_{1}$ are
given by%
\begin{align}
E_{10}^{1}  & \equiv M_{10}^{1+}M_{10}^{1}=s_{0}^{1}\bar{s}_{0}^{1}a_{0}%
^{1}\bar{a}_{0}^{1},\nonumber\\
E_{10}^{2}  & \equiv M_{10}^{2+}M_{10}^{2}=s_{0}^{2}\bar{s}_{0}^{2}a_{0}%
^{1}\bar{a}_{0}^{1}.
\end{align}
In this particular case, these POVMs satisfy the relations $E_{10}^{i}%
E_{10}^{j}=\delta_{ij}E_{10}^{i}$ (no sum over $i$) and%
\begin{equation}
E_{10}^{1}+E_{10}^{2}=I_{0}^{eff}.
\end{equation}
From these we find the conditional outcome rates%
\begin{equation}
\Pr(A_{1}^{1}|\Psi_{0})=|\alpha|^{2},\ \ \ \Pr(A_{1}^{2}|\Psi_{0})=|\beta
|^{2},
\end{equation}
assuming complete efficiency.

\section{The non-polarizing beamsplitter}

In this example, two beams of light enter the device through the two in-ports
and are passed on to two outcome detectors, as shown in Figure $2$.

\begin{figure}[h]
\centerline{\psfig{file=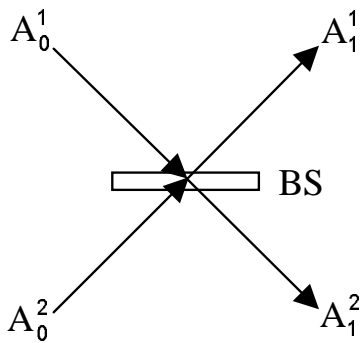,width=2in}}
\caption{The Beam-splitter.}%
\end{figure}

Typically, a single photon scenario is involved, which we discuss here.
For a single photon scenario, the initial state is given by%
\begin{align}
\Psi_{0}  & =\alpha|s^{1},0\rangle\otimes\mathbb{A}_{1,0}^{+}|0,0)+\beta
|s^{1},0\rangle\otimes\mathbb{A}_{2,0}^{+}|0,0)\nonumber\\
& =\alpha s_{0}^{1}a_{0}^{1}+\beta s^{1}a_{0}^{2},
\end{align}
where the same photon polarization state $s^{1}$ is assumed for both in-ports.
In the situation that different in-port states have different polarizations,
the dynamics is easily modified. For the case concerned, the dynamics is given
by the rules%
\begin{align}
U_{10}s_{0}^{1}a_{0}^{1}  & =ts_{1}^{1}a_{1}^{1}+irs_{1}^{1}a_{1}%
^{2},\nonumber\\
U_{10}s_{0}^{1}a_{0}^{2}  & =irs_{1}^{1}a_{1}^{1}+ts_{1}^{1}a_{1}^{2},
\end{align}
where $t$ and $r$ are real and satisfy $t^{2}+r^{2}=1$ and a relative phase
change associated with reflection has been taken into account. Using effective
completeness, we have%
\begin{equation}
U_{10}\backsimeq s_{1}^{1}\bar{s}_{0}^{1}\{ta_{1}^{1}+ira_{1}^{2}\}\bar{a}%
_{0}^{1}+s_{1}^{1}\bar{s}_{0}^{1}\{ira_{1}^{1}+ta_{1}^{2}\}\bar{a}_{0}^{2},
\end{equation}
which gives the non-zero POVM operators%
\begin{align}
E_{10}^{1}  & =s_{{0}}^{1}\bar{s}_{0}^{1}\{{r}^{2}a_{{0}}^{1}\bar{a}_{{0}}%
^{1}-irta_{{0}}^{1}\bar{a}_{{0}}^{2}+irta_{{0}}^{2}\bar{a}_{{0}}^{1}+{t}%
^{2}a_{{0}}^{2}\bar{a}_{{0}}^{2}\}\nonumber\\
E_{10}^{2}  & =s_{{0}}^{1}\bar{s}_{{0}}^{1}\{t^{2}a_{{0}}^{1}\bar{a}_{{0}}%
^{1}+irta_{{0}}^{1}\bar{a}_{{0}}^{2}-irta_{{0}}^{2}\bar{a}_{{0}}^{1}+{r}%
^{2}a_{{0}}^{2}\bar{a}_{{0}}^{2}\}.
\end{align}
This gives the non-zero outcome transition relative probability rates%
\begin{align}
\Pr(A_{1}^{1}|\Psi_{0})  & ={r}^{2}|\alpha|^{2}+irt(\alpha\beta^{\ast}%
-\alpha^{\ast}\beta)+{t}^{2}|\beta|^{2},\nonumber\\
\Pr(A_{1}^{2}|\Psi_{0})  & ={t}^{2}|\alpha|^{2}-irt(\alpha\beta^{\ast}%
-\alpha^{\ast}\beta)+{r}^{2}|\beta|^{2},
\end{align}
assuming complete efficiency and wave-train overlap. The computer algebra
program confirms that $\Pr(a_{1}^{0}|\Psi_{0})=0$ and $\Pr(A_{1}^{1}A_{1}%
^{2}|\Psi_{0})\equiv\Pr(a_{1}^{3}|\Psi_{0})=0$.

\section{Brandt's network}

The next example is a quantum optics network discussed by Brandt
\cite{BRANDT-1999} in terms of conventional POVMs and is shown in Figure $3$.
Here and elsewhere $M$ refers to a mirror.

\begin{figure}[h]
\centerline{\psfig{file=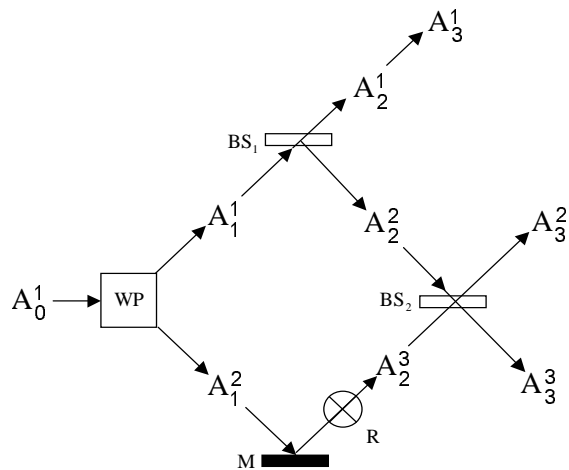,width=3in}}
\caption{Brandt's network.}%
\end{figure}


Brandt's analysis was in terms of non-orthogonal SUO state vectors. Our
analysis avoids non-orthogonality issues directly. The initial state is given
by
\begin{equation}
\Psi_{0}=(\alpha s_{0}^{1}+\beta s_{0}^{2})a_{0}^{1},
\end{equation}
where $s_{0}^{1},s_{0}^{2}$ denote orthogonal photon polarizations and
$\alpha$ and $\beta$ satisfy $|\alpha|^{2}+|\beta|^{2}=1$. Evolution from
$\Omega_{0}$ to $\Omega_{1}$ is given by the Wollaston prism dynamics given by
(\ref{POVM-02}). The transition from $\Omega_{1}$ to $\Omega_{2}$ is
determined by the effective transition rules%
\begin{align}
U_{21}s_{1}^{1}a_{1}^{1}  & =t_{1}s_{2}^{1}a_{2}^{1}+ir_{1}s_{2}^{1}a_{2}%
^{2},\nonumber\\
U_{21}s_{1}^{2}a_{1}^{2}  & =-s_{2}^{1}a_{2}^{4},\label{POVM-10}%
\end{align}
where $a_{2}^{4}\equiv\mathbb{A}_{3,2}^{+}|0,2)=|2^{3-1},2)=|4,2)$ and
$t_{1}^{2}+r_{1}^{2}=1$. The second equation in (\ref{POVM-10}) represents a
$\tfrac{1}{2}\pi$ rotation of the photon polarization vector as it passes
through the module labelled $R$ in Figure $3$. It is assumed that the
transmission and reflection parameters $t_{1}$, $r_{1}$ at $BS_{1}$ can be
arranged to have specific values, so these are left undeclared at this point
in the calculation. Hence%
\begin{equation}
U_{21}\backsimeq s_{2}^{1}\bar{s}_{1}^{1}(t_{1}a_{2}^{1}+ir_{1}a_{2}^{2}%
)\bar{a}_{1}^{1}-s_{2}^{1}\bar{s}_{1}^{2}a_{2}^{4}\bar{a}_{1}^{2}.
\end{equation}

In principle, detection can take place at $A_{2}^{1}$ during $\Omega_{2}$.
However, this can be placed $``$on hold$"$ until the final stage $\Omega_{3}$,
which is more convenient. This is represented by the rule%
\begin{equation}
U_{32}s_{2}^{1}a_{2}^{1}=s_{3}^{1}a_{3}^{1}.
\end{equation}
The other parts of the process from $\Omega_{2}$ to $\Omega_{3}$ involve
beam-splitter $BS_{2}$ and are given by%
\begin{align*}
U_{32}s_{2}^{1}a_{2}^{2}  & =t_{2}s_{3}^{1}a_{3}^{4}+ir_{2}s_{3}^{1}a_{3}%
^{2},\\
U_{32}s_{2}^{1}a_{2}^{4}  & =ir_{2}s_{3}^{1}a_{3}^{4}+t_{2}s_{3}^{1}a_{3}^{2},
\end{align*}
where $t_{2}^{2}+r_{2}^{2}=1$, giving%
\begin{align}
U_{32}  & \backsimeq s_{3}^{1}\bar{s}_{2}^{1}a_{3}^{1}\bar{a}_{2}^{1}%
+s_{3}^{1}\bar{s}_{2}^{1}(t_{2}a_{3}^{4}+ir_{2}a_{3}^{2})\bar{a}_{2}%
^{2}\nonumber\\
& \ \ \ \ \ +s_{3}^{1}\bar{s}_{2}^{1}(ir_{2}a_{3}^{4}+t_{2}a_{3}^{2})\bar
{a}_{2}^{4}.
\end{align}

The complete effective evolution operator $U_{31}\equiv U_{32}U_{21}U_{10}$
gives three non-zero POVMs:
\begin{align}
E_{30}^{1}  & ={t_{{1}}}^{2}s_{{0}}^{1}\bar{s}_{{0}}^{1}a_{{0}}^{1}\bar{a}%
_{0}^{1},\nonumber\\
E_{30}^{2}  & =\{{r_{{1}}}^{2}{r_{{2}}}^{2}s_{{0}}^{1}\bar{s}_{{0}}^{1}%
+r_{{1}}r_{{2}}t_{{2}}(s_{{0}}^{1}\bar{s}_{{0}}^{2}+s_{{0}}^{2}\bar{s}_{{0}%
}^{1})+{t_{{2}}}^{2}s_{{0}}^{2}\bar{s}_{{0}}^{2}\}a_{{0}}^{1}\bar{a}_{{0}}%
^{1}\nonumber\\
E_{30}^{4}  & =\{{t_{{2}}}^{2}{r_{{1}}}^{2}s_{{0}}^{1}\bar{s}_{{0}}^{1}%
-r_{{1}}r_{{2}}t_{{2}}(s_{{0}}^{1}\bar{s}_{{0}}^{2}+s_{{0}}^{2}\bar{s}_{{0}%
}^{1})+{r_{{2}}}^{2}s_{{0}}^{2}\bar{s}_{{0}}^{2}\}a_{{0}}^{1}\bar{a}_{{0}}%
^{1},\nonumber\\
&
\end{align}
which lead to the three non-zero outcome rates%
\begin{align}
\Pr(A_{2}^{1}|\Psi_{0})  & ={t_{{1}}}^{2}|\alpha|^{2},\nonumber\\
\Pr(A_{3}^{2}|\Psi_{0})  & ={r_{{1}}}^{2}{r_{{2}}}^{2}|\alpha|^{2}+r_{{2}%
}r_{{1}}t_{{2}}(\alpha^{\ast}\beta+\alpha\beta^{\ast})+{t_{{2}}}^{2}%
|\beta|^{2}\nonumber\\
\Pr(A_{3}^{3}|\Psi_{0})  & ={r_{{1}}}^{2}{t_{{2}}}^{2}|\alpha|^{2}-r_{{2}%
}r_{{1}}t_{{2}}(\alpha^{\ast}\beta+\alpha\beta^{\ast})+{r_{{2}}}^{2}%
|\beta|^{2},\nonumber\\
&
\end{align}
assuming perfect efficiency and wave-train overlap. When the reflection and
transmission coefficients are chosen as by Brandt \cite{BRANDT-1999}, these
rates agree with his precisely.

\section{The Franson-Bell experiment}

We now discuss the Franson-Bell experiment in detail. In the following, we
refer to $``$photons$"$ as if they were actual particles, because this is
convenient and provides an intuitive picture. However, the quantum dynamics
shows that such a picture can be misleading. A better interpretation of a
photon is simply as a click in a detector.

The basic experiment consists of a coherent pair of photons sent in opposite
directions towards a pair of separated Mach-Zehnder interferometers, as shown
in Figure $4$. Each photon passes through its own interferometer and depending
on path taken, can suffer a change in phase $\phi$ and a time delay $\Delta
T$. In the proposed experiment, this time delay is assumed the same for each
interferometer, but the phase changes $\phi_{1}$, $\phi_{2}$ associated with
the different interferometers can be altered at will.

\begin{figure}[h]
\centerline{\psfig{file=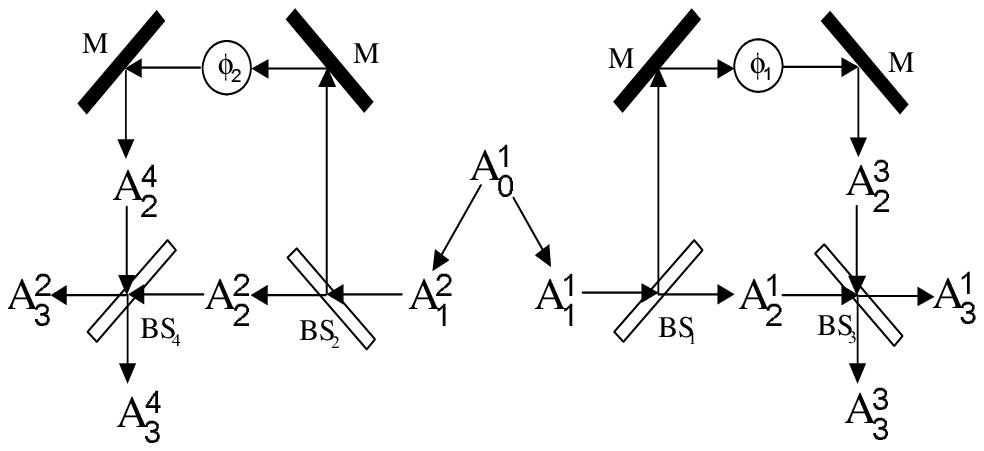,width=4in}}
\caption{The Franson-Bell experiment for $\Delta T<t_{2}\ll t_{1}$.}%
\end{figure}


The experiment hinges on the relationship between three characteristic times,
as discussed by Franson \cite{FRANSON-1989}. The first of these is the
coherence time of each photon. In real experiments, a $``$monochromatic$" $
single photon would be associated with a finite wave train of length $L$
moving at the speed of light $c$, which therefore takes a time $L/c$ to pass a
given point. In this experiment, the correct coherence timescale $t_{1}$ is
that associated with the production of the photon pair at $A_{0}^{1}$. This is
a characteristic of the photon pair source and of the collimation procedures
applied subsequently. Whilst $t_{1}$ cannot be altered, it can be determined
empirically. We assume $t_{1}$ is the same for each photon.

The second characteristic time is $t_{2}$, the effective time interval within
which both photons in a pair can be said to have been emitted. This can be
measured by coincidence measurements with detectors $A_{3}^{1}$ and $A_{3}%
^{2}$ with all beam-splitters removed. It is assumed $t_{2}$ can be determined
empirically and that $t_{2}\ll t_{1}$. This last inequality is crucial to the
Franson-Bell experiment because when this inequality holds, the observer has
no way of knowing when a photon pair was created during the relatively long
time interval $t_{1}$. It is this lack of knowledge which leads to quantum
interference in scenario $iii$) discussed below. The spectacular aspect of the
Franson Bell experiment is that unlike the double -slit experiment, where the
observer cannot know from which point in space a photon came, here the
observer does not know at which point in time the photon pair was produced.

The third characteristic time is $\Delta T$, the time difference between a
photon travelling along the short arm of its interferometer and along its long
arm. This time is adjustable and is assumed the same for each interferometer.

There are three scenarios we shall discuss: $i)\ \Delta T\ll t_{2}$, $ii)$
$t_{1}\ll\Delta T$ and $iii)\ t_{2}\ll\Delta T\ll t_{1}$. In Franson's
original analysis \cite{FRANSON-1989}, photon spin did not play a role.
Therefore, in all scenarios considered here, photon spin is assumed fixed once
a given photon pair has been created.

\section{Scenario $i)$ $\Delta T\ll t_{2}$}

The relevant figure for this scenario is Figure $4$.


The initial state is
\begin{equation}
\Psi_{0}=s_{0}A_{0}^{1},
\end{equation}
where $s_{0}$ represents the initial source spin state. The creation of a
photon pair is represented by the evolution operator
\begin{equation}
U_{1,0}s_{0}A_{0}^{1}=s_{1}A_{1}^{1}A_{1}^{2},
\end{equation}
where $s_{1}$ represent the combined spin state of the photon pair at stage
$\Omega_{1}$. Hence
\begin{equation}
U_{1,0}\backsimeq s_{1}\bar{s}_{0}a_{1}^{3}\bar{a}_{0}^{1}.
\end{equation}
Next,%
\begin{equation}
U_{2,1}s_{1}A_{1}^{1}A_{1}^{2}=s_{2}\{t_{1}A_{2}^{1}+ir_{1}e^{i\phi_{1}}%
A_{2}^{3}\}\{t_{2}A_{2}^{2}+ir_{2}e^{i\phi_{2}}A_{2}^{4}\},
\end{equation}
where $\phi_{1}$ and $\phi_{2}$ are total phase change factors due to the
increased path length of the long arms of the interferometers and phase-shift
plates introduced in those long arms by the observer. This gives%
\begin{align}
U_{2,1}  & \backsimeq s_{2}\bar{s}_{1}\{t_{1}t_{2}a_{2}^{3}+ir_{1}%
t_{2}e^{i\phi_{1}}a_{2}^{6}+it_{1}r_{2}e^{i\phi_{2}}a_{2}^{9}\nonumber\\
& \ \ \ \ \ \ \ \ \ \ \ \ \ \ \ \ \ \ \ \ -r_{1}r_{2}e^{i(\phi_{1}+\phi_{2}%
)}a_{2}^{12}\}\bar{a}_{1}^{3}.
\end{align}

There are four terms to consider in the transition from $\Omega_{2}$ to
$\Omega_{3}$:%
\begin{align}
U_{3,2}s_{2}A_{2}^{1}A_{2}^{2}  & =s_{3}\{t_{3}A_{3}^{1}+ir_{3}A_{3}%
^{3}\}\{t_{4}A_{3}^{2}+ir_{4}A_{3}^{4}\},\nonumber\\
U_{3,2}s_{2}A_{2}^{3}A_{2}^{2}  & =s_{3}\{t_{3}A_{3}^{3}+ir_{3}A_{3}%
^{1}\}\{t_{4}A_{3}^{2}+ir_{4}A_{3}^{4}\},\nonumber\\
U_{3,2}s_{2}A_{2}^{1}A_{2}^{4}  & =s_{3}\{t_{3}A_{3}^{1}+ir_{3}A_{3}%
^{3}\}\{t_{4}A_{3}^{4}+ir_{4}A_{3}^{2}\},\label{POVM-06}\\
U_{3,2}s_{2}A_{2}^{3}A_{2}^{4}  & =s_{3}\{t_{3}A_{3}^{3}+ir_{3}A_{3}%
^{1}\}\{t_{4}A_{3}^{4}+ir_{4}A_{3}^{2}\},\nonumber
\end{align}
which gives%
\begin{align}
U_{3,2}  & \backsimeq s_{3}\bar{s}_{2}\{t_{3}t_{4}a_{3}^{3}+ir_{3}t_{4}%
a_{3}^{6}+it_{3}r_{4}a_{3}^{9}-r_{3}r_{4}a_{3}^{12}\}\bar{a}_{2}%
^{3}\nonumber\\
& \ \ \ +\{t_{3}t_{4}a_{3}^{6}+ir_{3}t_{4}a_{3}^{3}+it_{3}r_{4}a_{3}%
^{12}-r_{3}r_{4}a_{3}^{9}\}\bar{a}_{2}^{6}\nonumber\\
& \ \ \ +\{t_{3}t_{4}a_{3}^{9}+ir_{3}t_{4}a_{3}^{12}+it_{3}r_{4}a_{3}%
^{3}-r_{3}r_{4}a_{3}^{6}\}\bar{a}_{2}^{9}\label{POVM-07}\\
& \ \ \ +\{t_{3}t_{4}a_{3}^{12}+ir_{3}t_{4}a_{3}^{9}+it_{3}r_{4}a_{3}%
^{6}-r_{3}r_{4}a_{3}^{3}\}\bar{a}_{2}^{12}.\nonumber
\end{align}
The total effective transition operator $U_{3,0}\equiv U_{3,2}U_{2,1}U_{1,0}$
was evaluated via computer algebra and from this four non-zero POVMs were
found. Combining these with the initial state and setting $t_{i}=r_{i}%
=1/\sqrt{2}$, $i=1,2,3,4$, as assumed by Franson \cite{FRANSON-1989}, gives
the relative coincidence rates%
\begin{align}
\Pr(A_{3}^{1}A_{3}^{2}|\Psi_{0})  & =\sin^{2}(\tfrac{1}{2}\phi_{1})\sin
^{2}(\tfrac{1}{2}\phi_{2}),\nonumber\\
\Pr(A_{3}^{3}A_{3}^{2}|\Psi_{0})  & =\cos^{2}(\tfrac{1}{2}\phi_{1})\sin
^{2}(\tfrac{1}{2}\phi_{2}),\nonumber\\
\Pr(A_{3}^{1}A_{3}^{4}|\Psi_{0})  & =\sin^{2}(\tfrac{1}{2}\phi_{1})\cos
^{2}(\tfrac{1}{2}\phi_{2}),\\
\Pr(A_{3}^{3}A_{3}^{4}|\Psi_{0})  & =\cos^{2}(\tfrac{1}{2}\phi_{1})\cos
^{2}(\tfrac{1}{2}\phi_{2}).\nonumber
\end{align}

Each of these rates shows angular dependence due to independent $``$photon
self-interference$"$ within each separate interferometer. This form of
interference will be referred to as \emph{local}. There are no \emph{global
}interference effects involving both interferometers and no post selection of
data is required.

\section{Scenario $ii)$ $t_{1}\ll\Delta T$}

In this situation, the photon wavetrains $A_{2}^{3}$, $A_{2}^{4}$ reflected at
$BS_{1}$ and $BS_{2}$ respectively travel along the long arms of their
respective interferometers at the speed of light or less, depending on the
medium through which they move. Therefore, these wavetrains arrive at $BS_{3}
$ and $BS_{4}$ long after the transmitted wavetrains $A_{2}^{1}$ and
$A_{2}^{2}$ have impinged on $BS_{3}$ and $BS_{4}$. In consequence, no local
or global interference can take place. In fact, the observer can now obtain
total information concerning the timing of each coincidence outcome in every
run of the experiment and know precisely what path was taken by each photon.

Under this circumstance, the four original detectors $A_{3}^{1}$, $A_{3}^{2}$,
$A_{3}^{3}$ and $A_{3}^{4}$ assumed in the previous section have to be
regarded as eight separate detectors, $A_{3}^{i}$, $i=1,2,\ldots,8$, as shown
in Figure $5$. The first four of these register photon clicks from short path
photons whilst the last four signal clicks from those that have traveled the
long paths. This information is specific to each photon and does not involve
any photon pairs. Note that the final stage quantum register involved in this
scenario and the next is 256 dimensional. However, our computer algebra
programme has no difficulty dealing with this because it works only within
effective Hilbert spaces of greatly reduced dimensions.

\begin{figure}[t]
\centerline{\psfig{file=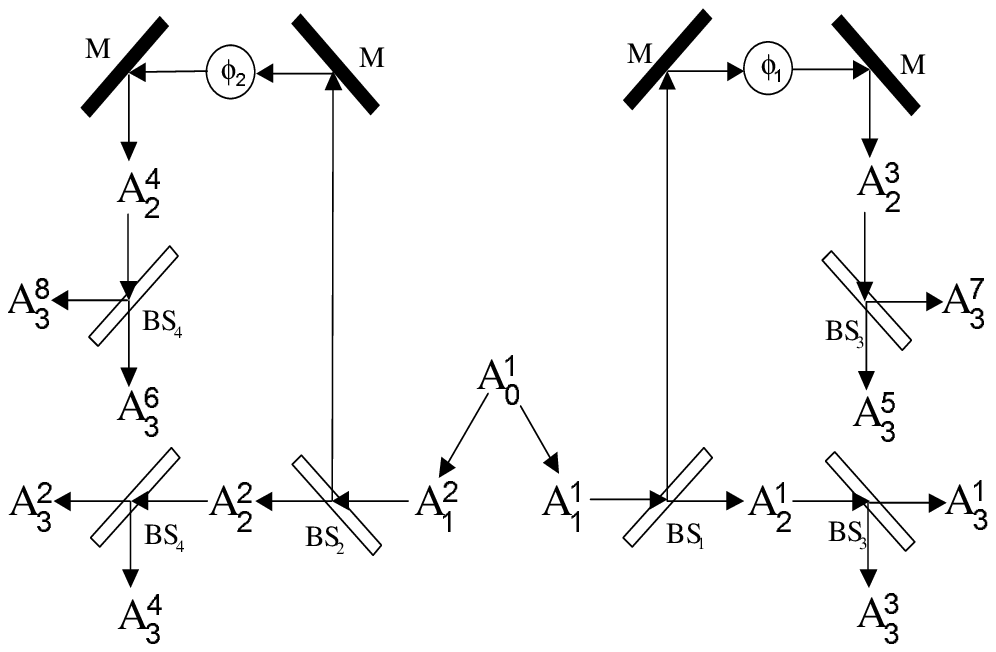,width=4in}}
\caption{The Franson-Bell experiment for $t_{2}\ll t_{1}\ll\Delta T$.}%
\end{figure}

This demonstrates a fundamental point about apparatus. In the conventional
usage of apparatus, experimentalists tend to regard their equipment as having
some sort of $``$trans-temporal$"$ identity, or persistence. In Figure $5$,
for example, beam-splitters $BS_{3}$ and $BS_{4}$ would most likely persist in
the laboratory during the long interval $\Delta T$ between their interaction
with wave-trains $A_{2}^{1}$ and $A_{2}^{2}$ and with the delayed wave-trains
$A_{2}^{3}$ and $A_{2}^{4}$. Even classically, however, this need not be the
case. It is conceivable that $\Delta T$ could be so long, such as several
years, so that the beam-splitters could be destroyed and rebuilt at leisure
between the observation of any short-arm photons and any long-arm photons.

Whatever the actuality in the laboratory, from a quantum point of view, the
beam splitters receiving short and long-arm photons should be considered as
completely separate pieces of equipment in this scenario (but not in the
next). In other words, apparatus and how it is used is time dependent. The
analysis in the next section shows that the rules for doing this can be quite
non-classical and appear to violate the ordinary rules of causality.

For the current scenario, $t_{1}\ll\Delta T$, the dynamics follows the same
rules as in the previous section up to the transition from $\Omega
_{2}\rightarrow\Omega_{3}$. At this point, the transformation rules have to
take into account the possibility that the observer could know the timings of
all events. The rules for this transition are now%
\begin{align}
U_{3,2}s_{2}A_{2}^{1}A_{2}^{2}  & =s_{3}\{t_{3}A_{3}^{1}+ir_{3}A_{3}%
^{3}\}\{t_{4}A_{3}^{2}+ir_{4}A_{3}^{4}\},\nonumber\\
U_{3,2}s_{2}A_{2}^{3}A_{2}^{2}  & =s_{3}\{t_{3}A_{3}^{5}+ir_{3}A_{3}%
^{7}\}\{t_{4}A_{3}^{2}+ir_{4}A_{3}^{4}\},\nonumber\\
U_{3,2}s_{2}A_{2}^{1}A_{2}^{4}  & =s_{3}\{t_{3}A_{3}^{1}+ir_{3}A_{3}%
^{3}\}\{t_{4}A_{3}^{6}+ir_{4}A_{3}^{8}\},\label{POVM-08}\\
U_{3,2}s_{2}A_{2}^{3}A_{2}^{4}  & =s_{3}\{t_{3}A_{3}^{5}+ir_{3}A_{3}%
^{7}\}\{t_{4}A_{3}^{6}+ir_{4}A_{3}^{8}\},\nonumber
\end{align}
which should be compared with (\ref{POVM-06}). This gives%
\begin{align}
U_{3,2}  & \backsimeq s_{3}\bar{s}_{2}\{t_{3}t_{4}a_{3}^{3}+ir_{3}t_{4}%
a_{3}^{6}+it_{3}r_{4}a_{3}^{9}-r_{3}r_{4}a_{3}^{12}\}\bar{a}_{2}%
^{3}\nonumber\\
& \ \ \ +\{t_{3}t_{4}a_{3}^{18}+ir_{3}t_{4}a_{3}^{24}+it_{3}r_{4}a_{3}%
^{66}-r_{3}r_{4}a_{3}^{72}\}\bar{a}_{2}^{6}\nonumber\\
& \ \ \ +\{t_{3}t_{4}a_{3}^{33}+ir_{3}t_{4}a_{3}^{36}+it_{3}r_{4}a_{3}%
^{129}-r_{3}r_{4}a_{3}^{132}\}\bar{a}_{2}^{9}\nonumber\\
& \ \ \ +\{t_{3}t_{4}a_{3}^{48}+ir_{3}t_{4}a_{3}^{96}+it_{3}r_{4}a_{3}%
^{144}-r_{3}r_{4}a_{3}^{192}\}\bar{a}_{2}^{12},\nonumber\\
& \label{POVM-11}%
\end{align}
which should be compared with (\ref{POVM-07}).

In this scenario, we find sixteen non-zero coincidence rates, each of the form
$\Pr(A_{3}^{i}A_{3}^{j}|\Psi_{0})$, where $i=1,3,5,7$ and $j=2,4,6,8.$ All of
them are constant, i.e., independent of $\phi_{1}$ and of $\phi_{2}$. For
example, $\Pr(A_{3}^{1}A_{3}^{6}|\Psi_{0})=t_{1}^{2}r_{2}^{2}t_{3}^{2}%
t_{4}^{2}$, and so on. In the case of symmetrical beam-splitters, where
$t_{i}=r_{i}=1/\sqrt{2}$, all sixteen rates are equal.

For this scenario, the detectors behave in a manner consistent with the notion
that photons are classical-like particles propagating along definite paths.

\section{Scenario $iii)$ $t_{2}\ll\Delta T\ll t_{1}$}

This is the scenario discussed by Franson \cite{FRANSON-1989}. In the
following, $\emph{S}$ stands for $``$short path$"$ and $L$ for $``$long path".
The fundamental change induced by the observer's setting of $\Delta T$ such
that $t_{2}\ll\Delta T\ll t_{1}$ is that unlike the previous scenario, the
observer cannot now use individual times of detector clicks to establish which
of the coincidences $S-S$ or $L-L$ has occurred in a given run of the
experiment. The relevant diagram is Figure $6$, which is identical to Figure
$5\ $except now $A_{3}^{5}$ is replaced by $A_{3}^{3\vee5},A_{3}^{7}$ replaced
by $A^{1\vee7}$, $A_{3}^{6}$ is replaced by $A_{3}^{4\vee6}$ and $A_{3}^{8}$
replaced by $A_{3}^{2\vee8}$, where for example $3\vee5$ means $``3$ or $5"$.
Which alternative is taken depends on the contextual information available in
principle to the observer.

\begin{figure}[t]
\centerline{\psfig{file=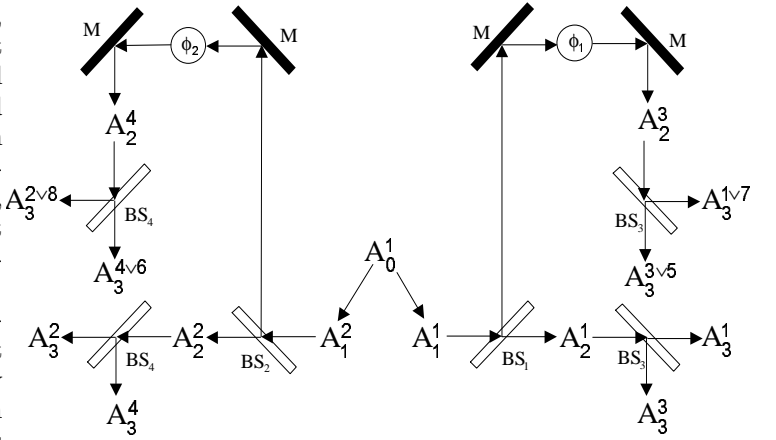,width=4in}}
\caption{The Franson-Bell experiment for $t_{2}\ll\Delta T\ll t_{1}$.}%
\end{figure}

The dynamics for this scenario is identical to that for the previous one,
except for the last equation in (\ref{POVM-08}), which is replaced by%
\begin{equation}
U_{3,2}s_{2}A_{2}^{3}A_{2}^{4}=s_{3}\{t_{3}A_{3}^{3}+ir_{3}A_{3}^{1}%
\}\{t_{4}A_{3}^{4}+ir_{4}A_{3}^{2}\}.
\end{equation}
This gives%
\begin{align}
U_{3,2}  & \backsimeq s_{3}\bar{s}_{2}\{t_{3}t_{4}a_{3}^{3}+ir_{3}t_{4}%
a_{3}^{6}+it_{3}r_{4}a_{3}^{9}-r_{3}r_{4}a_{3}^{12}\}\bar{a}_{2}%
^{3}\nonumber\\
& \ \ \ +\{t_{3}t_{4}a_{3}^{18}+ir_{3}t_{4}a_{3}^{24}+it_{3}r_{4}a_{3}%
^{66}-r_{3}r_{4}a_{3}^{72}\}\bar{a}_{2}^{6}\nonumber\\
& \ \ \ +\{t_{3}t_{4}a_{3}^{33}+ir_{3}t_{4}a_{3}^{36}+it_{3}r_{4}a_{3}%
^{129}-r_{3}r_{4}a_{3}^{132}\}\bar{a}_{2}^{9}\nonumber\\
& \ \ \ +\{t_{3}t_{4}a_{3}^{12}+ir_{3}t_{4}a_{3}^{9}+it_{3}r_{4}a_{3}%
^{6}-r_{3}r_{4}a_{3}^{3}\}\bar{a}_{2}^{12},\nonumber\\
&
\end{align}
instead of (\ref{POVM-11}).

Assuming a total production rate normalized to unity and symmetrical
beam-splitters, we find the coincidence rates%
\begin{align}
\Pr(A_{3}^{1}A_{3}^{2}|\Psi_{0})  & =\Pr(A_{3}^{3}A_{3}^{4}|\Psi_{0}%
)=\tfrac{1}{4}\cos^{2}(\frac{\phi_{1}+\phi_{2}}{2}),\nonumber\\
\Pr(A_{3}^{2}A_{3}^{3}|\Psi_{0})  & =\Pr(A_{3}^{1}A_{3}^{4}|\Psi_{0}%
)=\tfrac{1}{4}\sin^{2}(\frac{\phi_{1}+\phi_{2}}{2}),
\end{align}
which demonstrate non-locality. The other non-zero coincident rates involve
$A_{3}^{1}A_{3}^{6}$, $A_{3}^{1}A_{3}^{8},A_{3}^{2}A_{3}^{5},A_{3}^{2}%
A_{3}^{7},A_{3}^{3}A_{3}^{6},A_{3}^{3}A_{3}^{8}$, $A_{3}^{4}A_{3}^{5}$ and
$A_{3}^{4}A_{3}^{7}$ and are all $\frac{1}{16}$. Note that in actual Scenario
$iii$ experiments, the observer would have to measure the times at which
coincidence clicks were obtained and then post-select, i.e., filter out, those
coincidences corresponding to the $\{S-S$, $L-L\}$ processes and those
corresponding to $\{S-L,L-S\}$.

\section{Commentary and conclusions}

Since Franson's original paper, there has been great interest in empirical
confirmation of the Scenario $iii)$ predictions. Whilst there still appears
some room for debate concerning the interpretation of the experiment, the
results of Kwiat et al. \cite{KWIAT-1993} vindicate Franson's prediction,
which corresponds to $\Pr(A_{3}^{1}A_{3}^{2}|\Psi_{0})=%
\frac14
\cos^{2}(\frac{\phi_{1}+\phi_{2}}{2})$ in our approach.

Assuming the quantum theoretical interpretation of this experiment is correct,
then there is an extraordinary lesson to be learnt, not about SUOs in
particular, but about the rules concerning the use of apparatus and how these
can differ spectacularly to those expected classically. The interference of
the $S-S$ and $L-L$ amplitudes in Scenario $iii)$ cannot be envisaged in a
classical way to occur locally in time. Any attempt to think about it in terms
of photons as actual particles seems to lead to bizarre concepts which would
never be acceptable conventionally. It has to be recognized that a two-photon
state is not equivalent under all circumstances to a state with two separate
photons. A more recent quantum optics experiment with similar conclusions has
been reported by Kim \cite{KIM-2003}.

The weight of evidence points to the conclusion that quantum outcome
amplitudes are dynamically affected by \emph{contextual information} held in
principle by the observer. When some information is absent, then quantum
interference can occur. This supports the position of Heisenberg and Bohr
concerning the fundamental principles and interpretation of quantum physics.
Quantum optics experiments such as the Franson-Bell experiment are providing
more and more evidence that quantum mechanics is not just a theory of SUOs,
but also a fundamental perspective on the laws of observation in physics. It
is our view that the surface of those laws has only been scratched to date.
The formalism we have developed and presented here, coupled with modern
computer algebra technology, appears to give us some potential to dig deeper
into those laws.

\end{document}